\documentclass{moriond}

\usepackage{cite}
\usepackage{amsmath}
\usepackage{slashed}
\def\be{\begin{equation}}
\def\ee{\end{equation}}
\def\bea{\begin{eqnarray}}
\def\eea{\end{eqnarray}}

\newcommand{\Photo}{}

\begin{document}
\vspace*{4cm}
\title{LCSR predictions for $B \to K$ Hadronic Matrix Elements}

\author{ Dayanand Mishra$^{1,}$ }

\address{$^1$Universit\'e Claude Bernard Lyon 1, CNRS/IN2P3, \\
Institut de Physique des 2 Infinis de Lyon, UMR 5822, F-69622, Villeurbanne, France}

\maketitle\abstracts{
In this talk, I will present the calculation of LCSR predictions for the $B \to K$ Hadronic Matrix Elements (HME) at low $q^2$ using light meson distribution amplitudes. I will discuss the results obtained.}

\section{Introduction}

Decays of $B$ mesons into semi-leptonic final states mediated by neutral currents, such as $B \to K^{(*)}\ell\ell$, provide a valuable testing ground for the Standard Model (SM) of particle physics. These modes are loop suppressed in the SM and are theoretically clean, making them especially interesting for both theoretical and experimental studies.
Recent measurements of lepton flavor universality (LFU) ratios \cite{LHCb:2016due,LHCb:2022qnv,LHCb:2022vje,CMS:2024syx} show good agreement with SM predictions \cite{Hiller:2003js,Bordone:2016gaq,Mishra:2020orb}. However, it remains unclear whether new physics could still be affecting these decays in ways not captured by LFU observables.

One of the main challenges in this mode is the branching ratio, both total and differential. For instance, the total branching ratio measured by CMS and LHCb is $\text{BR}(B \to K \mu \mu) = (1.242 \pm 0.068) \times 10^{-7}$ \cite{CMS:2024syx} and $(1.186 \pm 0.034) \times 10^{-7}$ \cite{LHCb:2014cxe}, respectively, while theoretical predictions using form factors from lattice (HPQCD \cite{Parrott:2022zte}) and LCSRs (KR \cite{Khodjamirian:2017fxg}, GvDV \cite{Gubernari:2022hxn}) lie significantly higher, around $(1.9$–$2.3)\times 10^{-7}$. And for a comparison of differential branching ratios, see \cite{Mahmoudi:2024zna}. These differences highlight the need for more precise computations of the relevant hadronic matrix elements.

% One of the main challenges in this mode is the branching ratio, both differential and total. Significant discrepancies exist among different predictions and measurements. For example, the total branching ratios are: $\text{BR}(B \to K \mu \mu)|_{\text{CMS}} = (1.242 \pm 0.068) \times 10^{-7}$\cite{CMS:2024syx}, $\text{BR}(B \to K \mu \mu)|_{\text{LHCb}} = (1.186 \pm 0.034) \times 10^{-7}  $\cite{LHCb:2014cxe}, $\text{BR}(B \to K \mu \mu)|_{\text{HPQCD}} = (1.91 \pm 0.19) \times 10^{-7}$\cite{Parrott:2022zte}, $\text{BR}(B \to K \mu \mu)|_{\text{KR}} = (2.19 \pm 0.33) \times 10^{-7} 
% $\cite{Khodjamirian:2017fxg}, $\text{BR}(B \to K \mu \mu)|_{\text{GvDV}} = (2.3 \pm 0.2) \times 10^{-7}$\cite{Gubernari:2022hxn}. For a review of the differential branching ratio, see \cite{Mahmoudi:2024zna}. These differences highlight the need for more precise computations of the relevant hadronic matrix elements.

In this work, we focus on the computation of soft-gluon emission from the charm-loop. This effect has been previously studied by KMW \cite{Khodjamirian:2012rm} and GvDV \cite{Gubernari:2020eft} using $B$-meson distribution amplitudes (DAs). Here, we compute this contribution using light-meson DAs. There are two main motivations for this approach: first, light-meson DAs are better known than those of the $B$ meson; second, although this leads to a different light-cone operator product expansion (LCOPE), it describes the same physical process and offers a complementary perspective on the calculation.

\section{Undestanding Charm-loop}

\begin{figure}
\begin{minipage}{0.50\linewidth}
\centerline{\includegraphics[width=0.5\linewidth]{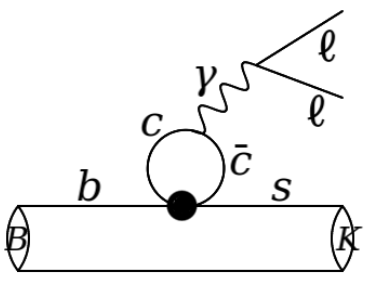}}
\label{charm_loop_sm}
\end{minipage}
\hfill
\begin{minipage}{0.50\linewidth}
\centerline{\includegraphics[width=0.5\linewidth]{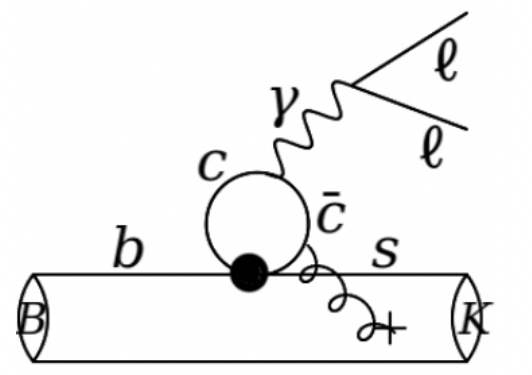}}
\label{charm_loop_sg}
\end{minipage}
\caption[]{(left) Short-distance charm-loop contribution to $B \to K\ell\ell$. (right) Non-factorizable soft-gluon emission from the charm loop in $B \to K\ell\ell$.
}
\label{charm_loop}
\end{figure}

In the SM, the relevant quark-level effective Hamiltonian \cite{Buchalla:1995vs}, with $\lambda_i = V_{ib}V_{is}^*$, is given by:
\begin{eqnarray}
    H_{\text{eff}} = \frac{4G_F}{\sqrt{2}}\left(\lambda_c \sum_{i=1,2} C_i \mathcal{O}_i + \lambda_t \sum_{i=3}^{10} \mathcal{C}_i \mathcal{O}_i\right),
\end{eqnarray}
where the relevant operators are:
\begin{eqnarray}
    \mathcal{O}_1 &=& (\bar{s}_i \gamma_\mu P_L c_j)(\bar{c}_j \gamma^\mu P_L b_i),  \,\,\;
    \mathcal{O}_2 = (\bar{s}_i \gamma_\mu P_L c_i)(\bar{c}_j \gamma^\mu P_L b_j), \nonumber\\
    \mathcal{O}_{7\gamma} &=& -\frac{e m_b}{16\pi^2}(\bar{s} \sigma_{\mu\nu} P_R b) F^{\mu\nu},  \,\,\;
    \mathcal{O}_9 = \frac{\alpha_{\text{em}}}{4\pi}(\bar{s} \gamma_\mu P_L b)(\bar{\ell} \gamma^\mu \ell), \nonumber\\
    \mathcal{O}_{10} &=& \frac{\alpha_{\text{em}}}{4\pi}(\bar{s} \gamma_\mu P_L b)(\bar{\ell} \gamma^\mu \gamma_5 \ell),
\end{eqnarray}
where $i, j$ are color indices and $e = \sqrt{4\pi \alpha_{\text{em}}}$. The operators $\mathcal{O}_{7\gamma}$, $\mathcal{O}_9$, and $\mathcal{O}_{10}$ arise from photon and $Z$ penguin diagrams, and box diagrams.
The quark-loop contribution, primarily from the current-current operators $\mathcal{O}_{1,2}$, is calculated perturbatively, including hard-gluon QCD corrections. This contribution is absorbed into the effective coefficient $C_9^{\text{eff}}$.
Due to the CKM structure and the relatively large charm quark mass, the dominant contribution comes from the charm-loop, as shown in Fig.~\ref{charm_loop}(a).
Contributions from light quarks are either CKM suppressed or associated with small Wilson coefficients and are thus not considered here.
In the SM, the Wilson coefficients at the scale relevant to $B$ decays are given by:
\begin{align}
C_1 = -0.2507, \quad  & C_2 = 1.0136,
\quad & C_{7\gamma} = -0.3143, \quad & C_9 = 4.0459, \quad & C_{10} = -4.2939,\nonumber
\end{align}
as reported in \cite{Bobeth:1999mk,Bobeth:2003at,Misiak:2004ew,Mahmoudi:2024zna}. The Wilson coefficients for $\mathcal{O}_{3-6}$ and $\mathcal{O}_{8g}$ are small and thus neglected in the discussion, although they can be included in a similar manner if needed.

Charm quarks also contribute through so-called ``charming penguins,'' where a $c\bar{c}$ pair forms charmonium resonances near $q^2 \sim 9$~GeV$^2$, leading to peaks in modes like $B \to J/\psi(\to \ell\ell) K$. These long-distance effects are much larger than the short-distance contributions and are non-perturbative, making them hard to compute reliably. To avoid this region, experiments impose cuts on the dilepton invariant mass, typically focusing on the low-$q^2$ window, $1 < q^2 < 6$~GeV$^2$, where theoretical predictions are more robust and LFU tests are cleaner.
Beyond resonances, charm loops also induce soft-gluon emissions with virtuality $|k^2| \ll 4m_c^2$, as shown in Fig.~\ref{charm_loop}(b). This non-perturbative effect, often called the charm-loop contribution, can introduce significant uncertainty at low-$q^2$. After Fierz rearrangement, the operators $C_1 \mathcal{O}_1 + C_2 \mathcal{O}_2$ split into color-singlet and color-octet structures, contributing respectively to factorizable and non-factorizable amplitudes. Care must be taken to avoid double-counting when modeling charm contributions across different categories such as loops, rescattering, and resonances.

The amplitude for the process is written as:
\begin{eqnarray}
        \langle K \ell\ell | H_{eff} |B\rangle &= \frac{\alpha}{4\pi} \frac{4G_F}{\sqrt{2}} V_{cb} V_{cs}^* \Bigg[(C_9 L^\mu_V + C_{10} L^\mu_A) \langle K|\bar{s}\gamma_\mu P_L b|B\rangle-\frac{16\pi^2}{q^2} L^\mu_V \langle K|\mathcal{H}_\mu |B\rangle \Big]
\end{eqnarray}
with,
\begin{eqnarray}
     L^\mu_{V(A)}&=&\bar{\ell}\gamma^\mu (\gamma_5) \ell, \ \text{and}\, 
    \mathcal{H}_\mu =i\int d^4x e^{iq.x}T\lbrace{ j_\mu^{em}(x),\left(C_1+\frac{C_2}{3}\right)\mathcal{O} + 2C_2 \tilde{\mathcal{O}}\rbrace},\nonumber
\end{eqnarray}
   
\section{Non-factorizable charm-loop}

The non-factorizable charm-loop contribution arises from soft-gluon emission off the charm loop, with gluon virtuality $|k^2| \ll 4m_c^2$, as shown in Fig.~\ref{charm_loop}. This effect goes beyond perturbation theory and can be a major source of uncertainty at low $q^2$.
The relevant correlator is:
\begin{equation}
  \mathcal{H}_{\mu,\,\text{non-fac}} \sim \int d^4x\, e^{iq\cdot x} T\left\{ j_\mu^{\text{em}}(x), \tilde{\mathcal{O}} \right\}, \label{charmloopeq1}
\end{equation}
which shows light-cone dominance for $q^2 \ll 4m_c^2$ \cite{Khodjamirian:2010vf}. Applying field redefinitions for the charm fields, one finds that the dominant contributions come from regions where $x^2 \approx 0$.
Using the gluon-field-strength part of the charm propagator and performing a light-cone expansion, the non-factorizable contribution to the hadronic matrix element is expressed as:
\begin{align}
    \langle K| \mathcal{H}_{\mu,\,\text{non-fac}} | B \rangle_{tw-3} \propto \int d\omega\, d\alpha_s\, d\alpha_d \, \phi_{3K}(\alpha_i,\mu)\, \tilde{I}_{\mu\rho\alpha\beta}(q,\omega)\, \cdots
\end{align}
where $\tilde{I}_{\mu\rho\alpha\beta}$ is a perturbatively calculable kernel (see \cite{Mahajan:2024xpo} for detailed study). However, symmetry arguments—particularly contractions involving the Levi-Civita tensor—cause the entire contribution to vanish at twist-3. A similar calculation shows that the twist-4 term also vanishes:
\begin{equation}
    \langle K| \mathcal{H}_{\mu,\,\text{non-fac}} | B \rangle_{\text{twist-3 + twist-4}} = 0.
\end{equation}

This is the main result of our analysis. Unlike the case using $B$-meson DAs, where the contribution was small but nonzero, the use of kaon DAs leads to an exact cancellation at leading twist levels. This makes the computation simpler. Employing local quark-hadron duality (QHD), the full amplitude vanishes. 
We also analyzed $B \to K^*$ decays and found that the twist-3 contribution vanishes:
\begin{equation}
    \langle K^*| \mathcal{H}_{\mu,\,\text{non-fac}} | B \rangle_{twist-3} = 0.
\end{equation}
While both the light-meson and $B$-meson DA approaches rely on different light-cone expansions, their convergence implies that comparing their results at a given twist level is meaningful. Both approaches consistently show that non-factorizable charm-loop effects are small, reinforcing the reliability of theoretical predictions in the low-$q^2$ region.

We also explored the computation of local HME directly, avoiding the use of semi-global quark-hadron duality. Preliminary results indicate that such a formulation, currently under investigation, can offer a more controlled and potentially more reliable alternative to QHD-based estimates\cite{Carvunis:2024koh,carvunis}.

\section{Summary and Discussion}

Even in the theoretically clean low-$q^2$ region of $B \to K^{(*)} \ell\ell$ decays, uncertainties beyond form factors—particularly from non-factorizable charm-loop effects—remain important. Previous estimates based on $B$-meson LCDAs found these contributions to be small \cite{Khodjamirian:2010vf,Gubernari:2020eft}.

In this work, we independently revisited these contributions using light-meson LCDAs. We found that the non-factorizable charm-loop effects vanish up to twist-4 for $B \to K \ell\ell$, and similarly for $B \to K^* \ell\ell$. This result arises naturally from symmetry arguments—specifically, the contraction of antisymmetric Levi-Civita structures with symmetric light-meson distribution amplitudes.
These findings support the conclusion that non-factorizable charm-loop contributions are negligible.

% \section*{Acknowledgments}
% % I Thank Namit Mahajan for 

\section*{References}
\bibliography{Dayanand}

\begin{thebibliography}{10}

\bibitem{LHCb:2016due}
R.~Aaij et~al.
\newblock {Measurement of the phase difference between short- and long-distance
  amplitudes in the $B^{+}\to K^{+}\mu^{+}\mu^{-}$ decay}.
\newblock {\em Eur. Phys. J. C}, 77(3):161, 2017.

\bibitem{LHCb:2022qnv}
R.~Aaij et~al.
\newblock {Test of lepton universality in $b \rightarrow s \ell^+ \ell^-$
  decays}.
\newblock {\em Phys. Rev. Lett.}, 131(5):051803, 2023.

\bibitem{LHCb:2022vje}
R.~Aaij et~al.
\newblock {Measurement of lepton universality parameters in $B^+\to
  K^+\ell^+\ell^-$ and $B^0\to K^{*0}\ell^+\ell^-$ decays}.
\newblock {\em Phys. Rev. D}, 108(3):032002, 2023.

\bibitem{CMS:2024syx}
A.~Hayrapetyan et~al.
\newblock {Test of lepton flavor universality in $B^{\pm}\to K^{\pm}\mu^+\mu^-$
  and $B^{\pm}\to K^{\pm}e^+e^-$ decays in proton-proton collisions at
  $\sqrt{s}$ = 13 TeV}.
\newblock {\em Rept. Prog. Phys.}, 87(7):077802, 2024.

\bibitem{Hiller:2003js}
G.~Hiller and F.~Kruger.
\newblock {More model-independent analysis of $b \to s$ processes}.
\newblock {\em Phys. Rev. D}, 69:074020, 2004.

\bibitem{Bordone:2016gaq}
M.~Bordone, G.~Isidori, and A.~Pattori.
\newblock {On the Standard Model predictions for $R_K$ and $R_{K^*}$}.
\newblock {\em Eur. Phys. J. C}, 76(8):440, 2016.

\bibitem{Mishra:2020orb}
D.~Mishra and N.~Mahajan.
\newblock {Impact of soft photons on $B\rightarrow K \ell^+ \ell^-$}.
\newblock {\em Phys. Rev. D}, 103(5):056022, 2021.

\bibitem{LHCb:2014cxe}
R.~Aaij et~al.
\newblock {Differential branching fractions and isospin asymmetries of $B \to
  K^{(*)} \mu^+ \mu^-$ decays}.
\newblock {\em JHEP}, 06:133, 2014.

\bibitem{Parrott:2022zte}
W.~G. Parrott, C.~Bouchard, and C.~T.~H. Davies.
\newblock {Standard Model predictions for $B\to
  K\ensuremath{\ell}^+\ensuremath{\ell}^-, B\to K\ensuremath{\ell}_1^-
  \ensuremath{\ell}_2^+ { \rm and } B\to
  K\ensuremath{\nu}\ensuremath{\bar{\nu}}\textasciimacron{}$ using form factors
  from $N_f=2+1+1$ lattice QCD}.
\newblock {\em Phys. Rev. D}, 107(1):014511, 2023.
\newblock [Erratum: Phys.Rev.D 107, 119903 (2023)].

\bibitem{Khodjamirian:2017fxg}
A.~Khodjamirian and A.~V. Rusov.
\newblock {$B_{s}\to K \ell \nu_\ell$ and $B_{(s)} \to \pi (K) \ell^+\ell^-$
  decays at large recoil and CKM matrix elements}.
\newblock {\em JHEP}, 08:112, 2017.

\bibitem{Gubernari:2022hxn}
N.~Gubernari, M.~Reboud, D.~van Dyk, and J.~Virto.
\newblock {Improved theory predictions and global analysis of exclusive $b \to
  s\mu^+\mu^-$ processes}.
\newblock {\em JHEP}, 09:133, 2022.

\bibitem{Mahmoudi:2024zna}
F.~Mahmoudi and Y.~Monceaux.
\newblock {Overview of $B \to K^{(*)}\ell \ell$ Theoretical Calculations and
  Uncertainties}.
\newblock {\em Symmetry}, 16(8):1006, 2024.

\bibitem{Khodjamirian:2012rm}
A.~Khodjamirian, Th. Mannel, and Y.~M. Wang.
\newblock {$B \to K \ell^{+}\ell^{-}$ decay at large hadronic recoil}.
\newblock {\em JHEP}, 02:010, 2013.

\bibitem{Gubernari:2020eft}
N.~Gubernari, D.~van Dyk, and J.~Virto.
\newblock {Non-local matrix elements in $B_{(s)}\to
  \{K^{(*)},\phi\}\ell^+\ell^-$}.
\newblock {\em JHEP}, 02:088, 2021.

\bibitem{Buchalla:1995vs}
G.~Buchalla, A.~J. Buras, and M.~E. Lautenbacher.
\newblock {Weak decays beyond leading logarithms}.
\newblock {\em Rev. Mod. Phys.}, 68:1125--1144, 1996.

\bibitem{Bobeth:1999mk}
C.~Bobeth, M.~Misiak, and J.~Urban.
\newblock {Photonic penguins at two loops and $m_t$ dependence of $BR[B \to X_s
  l^+ l^-]$}.
\newblock {\em Nucl. Phys. B}, 574:291--330, 2000.

\bibitem{Bobeth:2003at}
C.~Bobeth, P.~Gambino, M.~Gorbahn, and U.~Haisch.
\newblock {Complete NNLO QCD analysis of $\bar{B} \to X_{(s)} l^+ l^-$ and
  higher order electroweak effects}.
\newblock {\em JHEP}, 04:071, 2004.

\bibitem{Misiak:2004ew}
M.~Misiak and M.~Steinhauser.
\newblock {Three loop matching of the dipole operators for $b \to s \gamma$ and
  $b \to s g$}.
\newblock {\em Nucl. Phys. B}, 683:277--305, 2004.

\bibitem{Khodjamirian:2010vf}
A.~Khodjamirian, Th. Mannel, A.~A. Pivovarov, and Y.~M. Wang.
\newblock {Charm-loop effect in $B \to K^{(*)} \ell^{+} \ell^{-}$ and $B\to
  K^*\gamma$}.
\newblock {\em JHEP}, 09:089, 2010.

\bibitem{Mahajan:2024xpo}
N.~Mahajan and D.~Mishra.
\newblock {Smallness of charm-loop effects in $B\to
  K^{(*)}\ensuremath{\ell}\ensuremath{\ell}$ at low $q^2$: Light-meson
  distribution-amplitude analysis}.
\newblock {\em Phys. Rev. D}, 111(3):L031504, 2025.

\bibitem{Carvunis:2024koh}
A.~Carvunis, F.~Mahmoudi, and Y.~Monceaux.
\newblock {Potential of light-cone sum rules without semiglobal quark-hadron
  duality}.
\newblock {\em Phys. Rev. D}, 110(11):114008, 2024.

\bibitem{carvunis}
A.~Carvunis, F.~Mahmoudi, D.~Mishra, and Y.~Monceaux.
\newblock {Calculation of Local $B \to K$ Form-factors using light meson DAs}.
\newblock {\em In preparation}.

\end{thebibliography}

%%% manually generated bibliography
%\begin{thebibliography}{99}
%\bibitem{ja}C Jarlskog in {\em CP Violation}, ed. C Jarlskog
%(World Scientific, Singapore, 1988).
%\bibitem{ma}L. Maiani, \Journal{\PLB}{62}{183}{1976}.
%\bibitem{bu}J.D. Bjorken and I. Dunietz, \Journal{\PRD}{36}{2109}{1987}.
%\bibitem{bd}C.D. Buchanan {\it et al}, \Journal{\PRD}{45}{4088}{1992}.
%\end{thebibliography}

\end{document}